\documentstyle[12pt,graphicx,caption2]{article}

\begin{document}

\title{New model of the kaon regeneration}
\author{V.I. Nazaruk\\
Institute for Nuclear Research of RAS, 60th October\\
Anniversary Prospect 7a, 117312 Moscow, Russia.*}

\date{}
\maketitle
\bigskip

\begin{abstract}
It is shown that in the standard model of $K^0_{S}$ regeneration a system of noncoupled equations of motion is used instead of the coupled ones. A model alternative to the standard one is proposed. The calculation performed by means of diagram technique agrees with that based on exact solution of equations of motion.

\end{abstract}

\vspace{5mm}
{\bf PACS:} 11.30.Fs; 13.75.Cs

\vspace{5mm}
Keywords: equations of motion, regeneration, diagram technique  

\vspace{1cm}

*E-mail: nazaruk@inr.ru

\newpage
\setcounter{equation}{0}
\section{Introduction}
The effect of kaon regeneration has been known since the 1950s. However, in the previous calculations [1-3] a system of noncoupled equations of motion was considered (see Eq. (1)) instead of the coupled ones. This is a fundamental defect because it leads to a qualitative disagreement in the results. This means that the regeneration has not been described at all. The result obtained in [2] was adduced in [4,5] and subsequent papers. In this paper we consider the model based on the exact solution of the coupled equations of motion with the potentials taken in  general form. The comparison with the previous model and our calculation performed by means of the diagram technique is given as well.

Let $K^0_{L}$ fall onto the plate at $t=0$. The probability of finding $K^0_S$ is of our particular interest. Our approach is as follows. Since $K^0N$- and $\bar{K}^0N$-interections are known, we go to $K^0,\bar{K}^0$ representation. The problem is described by coupled equations of motion for $K^0(t)$ and $\bar{K}^0(t)$. We find the corresponding solutions and revert to $K^0_{L},K^0_{S}$ representation. 

\section{Previous calculations}
In the previous calculations the starting equations are (see Eqs. (3) of Ref. [2]):
\begin{eqnarray}
(\partial_x-ink) \alpha =0,\nonumber\\
(\partial_x-in'k) \alpha '=0,
\end{eqnarray}
where $n$ and $n'$ are the indexes of refraction for $\alpha $ and $\alpha '$, respectively.
In this equation the change of variables $\alpha , \alpha '\rightarrow \alpha _1, \alpha _2$ is performed and effects of weak interactions are added. As a consequence of the change of variables, Eqs. (5) of [2] are coupled. The solution of (5) gives results (6) of Ref. [2] (or (1) of Ref. [3]). This result is adduced in Eq. (9.32) of Ref. [4], in Eqs. (7.83)-(7.89) of Ref. [5] and subsequent papers.

We consider all the possibilities. If $\alpha $ and $\alpha ' $ correspond to $K^0_{S}$ and $K^0_{L}$, $\alpha _1$ and $\alpha _2$ describe $K^0$ and $\bar{K}^0$, whereas our interest is with  $K^0_{S}$ and $K^0_{L}$. Besides, the indexes of refraction for $K^0_{L}$ 
and $K^0_{S}$ are unknown.

Let $\alpha = K^0$ and $\alpha '=\bar{K}^0$. Then $\alpha _1=K^0_{S}$ and $\alpha _2= K^0_{L}$. This variant follows from the initial conditions (see [2]): $\alpha _2(0)=1$, $\alpha _1(0)=0$. 
There is no off-diagonal mass $\epsilon =(m_L-m_S)/2$. Equations (1) are noncoupled. 
The noncoupled equations exist only for the stationary states and don't exist for $K^0$ and $\bar{K}^0$.

In any case Eqs. (1) are unrelated to the problem. Our calculation gives the inverse $\Delta \Gamma $- and $\Delta m$-dependences (see Eq. (24) of Ref. [6]). In this paper we consider the model with the potentials taken in general form. The similar question for $\Lambda \bar{\Lambda }$ oscillations is studied in [7-10].

\section{Our model}
Let $K^0_{L}$ fall onto the plate at $t=0$. We use the model described in paragraph 2 of introduction. In Ref. [6] the exact wave function $K_S(t)$ of $K^0_{S}$ has been calculated. The probability of finding $K^0_{S}$ or, what is the same, the probability of $K^0_{L}K^0_{S}$ transition is given by Eq. (12) of Ref. [6]:
\begin{equation}
\mid \!K_S(t)\!\mid ^2=\frac {1}{4}\mid V/p\mid ^2e^{{\rm Im}Vt+2{\rm Im}Mt}[e^{-{\rm Im}(pt)}+e^{{\rm Im}(pt)}-e^{i{\rm Re}(pt)}-e^{-i{\rm  
Re}(pt)}]
\end{equation}
This expression is exact; $\mid \!K_S(t=0)\!\mid ^2=0$. 

If ${\rm Im}V=0$, then ${\rm Im}p=0$ as well. In this case
\begin{equation}
\mid \!K_S(t)\!\mid ^2=\mid V/p\mid ^2e^{-(\Gamma _{K^0}^a+\Gamma ^d)t}\sin ^2({\rm Re}(pt)/2).
\end{equation}
This is a pure oscillation regime. Here regeneration by scattering takes place. The regeneration by absorption is described by ${\rm Im}V$.

As in [6] we put $p\approx V+2 \epsilon ^2/V$ ($\epsilon =\Delta m/2$, $\Delta m=m_L-m_S$, where $m_L$ and $m_S$ are the masses of stationary states); $\Gamma _{K^0}^d=\Gamma _{\bar{K}^0}^d=\Gamma ^d$, $m_{K^0}=m_{\bar{K}^0}=m$ ($m$ and $\Gamma ^d$ are the mass and width of decay of $K^0$, respectively). Let's denote $\Gamma _{K^0}^a$ and $\Gamma _{\bar{K}^0}^a$ as widths of absorption (not decay) of $K^0$ and ${\bar{K}^0}$, respectively. Then
\begin{equation}
2{\rm Im}M=-(\Gamma _{K^0}^a+\Gamma ^d),
\end{equation}
\begin{equation}
{\rm Re}V={\rm Re}U_{\bar{K}^0}-{\rm Re}U_{K^0},
\end{equation}
\begin{equation}
{\rm Im}V=-\frac{\Delta \Gamma }{2},
\end{equation}
where
\begin{equation}
\Delta \Gamma =\Gamma _{\bar{K}^0}^a-\Gamma _{K^0}^a.
\end{equation}

Equation (2) gives:
\begin{equation}
\mid \!K_S(t)\!\mid ^2=R\omega ,
\end{equation}
\begin{equation}
R=\frac {1}{4}\mid V/p\mid ^2e^{-(\Gamma _{K^0}^a+\Gamma ^d)t},
\end{equation}
\begin{equation}
\omega =e^{-\Gamma (K_L\rightarrow K_S)t}+e^{[\Gamma (K_L\rightarrow K_S)t-\Delta \Gamma ]t}-2e^{-\Delta \Gamma t/2}\cos({\rm Re}(pt)),
\end{equation}
where
\begin{equation}
\Gamma (K_L\rightarrow K_S)=\frac{\epsilon ^2}{\mid V\mid ^2}\Delta \Gamma ,
\end{equation}
\begin{equation}
{\rm Re}p\approx {\rm Re}V+2\epsilon ^2\frac{{\rm Re}V}{\mid V\mid ^2}.
\end{equation}
$\Gamma (K_L\rightarrow K_S)$ is the width of $K_LK_S$ transition (regeneration). The value $\Delta m$ is involved in $\Gamma (K_L\rightarrow K_S)$ and $\cos({\rm Re}(pt))$.

Let
\begin{equation}
\Delta \Gamma t\gg 1.
\end{equation}
In this case
\begin{equation}
\omega =e^{-\Gamma (K_L\rightarrow K_S)t}.
\end{equation}
The $t$-dependence is given by the exponential decay law. It is significant that ${\rm Re}V\ne 0$ in contrast to [6]. (Note that (14) is valid if $\Delta \Gamma >2\epsilon $.)

\section{Connection between the models based on the diagram technique and the exact solution}
The calculation presented above is cumbersome and formal. So verification is required. In [11] an approach based on the perturbation theory was proposed. Regeneration followed by the decay $K^0_L\rightarrow K^0_S\rightarrow \pi \pi $ was considered. The similar approach is used for the $n\bar{n}$ transition in a medium followed by annihilation (see Refs. [12-15]). The process amplitude $M(K^0_L\rightarrow K^0_S\rightarrow \pi \pi )$ is 
\begin{equation}
M(K^0_L\rightarrow K^0_S\rightarrow \pi \pi )=
\frac{\epsilon }{V}M_d(K^0_S\rightarrow \pi \pi).
\end{equation}
(See the second term of Eq. (23) of Ref. [11].) Here $M_d(K^0_S\rightarrow \pi \pi)$ is the in-medium amplitude of the decay $K^0_S\rightarrow \pi \pi$. The corresponding process width is
\begin{equation}
\Gamma (K^0_L\rightarrow K^0_S\rightarrow \pi \pi)=\frac{\epsilon ^2}{\mid V\mid ^2}\Gamma _d(K^0_S\rightarrow \pi \pi ),
\end{equation}
where $\Gamma _d(K^0_S\rightarrow \pi \pi )$ is the width of the decay $K^0_S\rightarrow \pi \pi $.

Consider now the connection between the models based on the diagram technique and the exact solution. In this case we write (16) in the form
\begin{equation}
\Gamma (K^0_L\rightarrow K^0_S\rightarrow \pi \pi)=\frac{\epsilon ^2}{\mid V\mid ^2}\Gamma _d(K^0_S\rightarrow \pi \pi )\frac{\Delta \Gamma }{\Delta \Gamma }=
\Gamma (K_L\rightarrow K_S)W,
\end{equation}
\begin{equation}
W=\frac{\Gamma _d(K^0_S\rightarrow \pi \pi )}{\Delta \Gamma },
\end{equation}
where $W$ is the probability of the $K^0_S$ decay in the channel 
$K^0_S\rightarrow \pi \pi $. The physical sense of (17) is obvious: the multistep process $K^0_L\rightarrow K^0_S\rightarrow \pi \pi $ involves the subprocess of $K_LK_S$ transition (regeneration). Equation (17) is verification of the models considered above.

Due to a strong absorption of $\bar{K}^0$ and zero momentum transfer in the $K^0\bar{K}^0$ transition vertex the description of competition between scattering and absorption is of particular importance. In this regard the diagram technique has some advantage over the model based on the equations of motion (see Refs. [14,15]).

\section{Limiting case and numerical results}
Let us consider the limiting case $t\rightarrow 0$. Expanding (10) to the terms $\sim t^2$ we have
\begin{equation}
\omega =[1-\cos({\rm Re}(pt))](2-\Delta \Gamma t)+\omega _1t^2/2,
\end{equation}
\begin{equation}
\omega _1=[\Gamma (K_L\rightarrow K_S)]^2+[\Gamma (K_L\rightarrow K_S)- \Delta \Gamma ]^2-\frac{(\Delta \Gamma )^2}{2}\cos({\rm Re}(pt)).
\end{equation}

Let ${\rm Re}V\ne 0$ and $\Delta \Gamma =0$. Then $\Gamma (K_L\rightarrow K_S)=\omega _1=0$ and $\mid \!K_S(t)\!\mid ^2$ coincides with (3). The opposite case when ${\rm Re}V=0$ is more interesting. Then
\begin{equation}
\omega =\omega _1t^2/2
\end{equation}
and
\begin{equation}
\mid \!K_S(t)\!\mid ^2=\frac{1}{8}\mid V/p\mid ^2e^{-(\Gamma _{K^0}^a+\Gamma ^d)t}t^2.
\end{equation}
Here regeneration by absorption takes place. Comparing (19) and (21) we see that ${\rm Re}V\ne 0$ violates $t^2$-dependence.

\begin{figure}[h]
%  \reflectbox{\includegraphics[height=.3\textheight]{golfer}}
  {\includegraphics[height=.25\textheight]{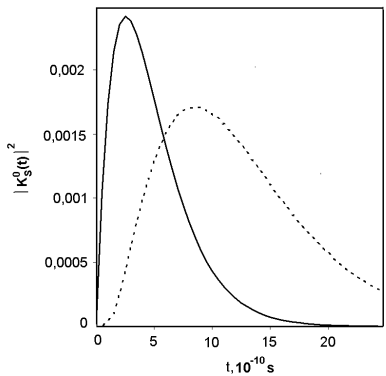}}
  \caption{ Probability of finding $K^0_S$. Solid and dashed curves correspond to ${\rm Re}V=\Delta \Gamma /2$ and calculation by means of Eq. (25), respectively.}
\end{figure}

\begin{figure}[h]
%  \reflectbox{\includegraphics[height=.3\textheight]{golfer}}
  {\includegraphics[height=.25\textheight]{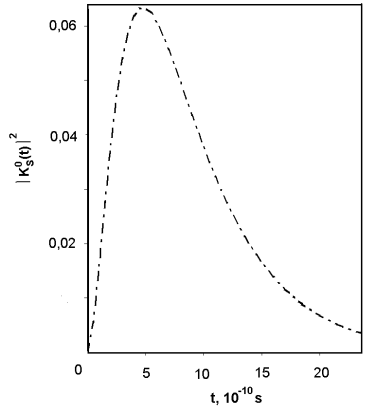}}
  \caption{ Probability of finding $K^0_S$. Dot-dashed curve corresponds to ${\rm Re}V=0$.}
\end{figure}

Let us revert to Eqs. (8)-(10). $\Gamma _{K^0}^a$ and $\Gamma _{\bar{K}^0}^a$ are given by standard expressions which follow from the optical theorem:
\begin{equation}
\Gamma _{K^0}^a=N_nv\sigma (K^0n)+N_pv\sigma (K^0p)
\end{equation}
and
\begin{equation}
\Gamma _{\bar{K}^0}^a=N_nv\sigma (\bar{K}^0n)+N_pv\sigma (\bar{K}^0p),
\end{equation}
where $N_n$ and $N_p$ are the number of netrons and protons in a unit of volume, respectively; $\sigma (K^0N)$ and $\sigma (\bar{K}^0N)$ are the total cross sections of $K^0N$- and $\bar{K}^0N$-interactions, $v$ is the velocity of the $K^0$ meson. By way of illustration we take $\sigma (K^0n)=\sigma (K^0p)=15$ mb [16]. As in Ref. [2], we use $\sigma (K^0N)=\frac{1}{3}\sigma (\bar{K}^0N)$.

Instead of cross sections one can use the forward scattering amplitudes of kaons by the molecules of the medium. In this case 
\begin{equation}
V=U_{\bar{K}^0}-U_{K^0}=\frac{2\pi }{m}N_mf_{21},
\end{equation}
$f_{21}=f-\bar{f}$. Here $N_m$ is the number of molecules in a unit of volume, $f$ and $\bar{f}$ are the forward scattering amplitudes of $K^0$ and $\bar{K}^0$, respectively.

For the copper absorber the probability of finding $K^0_{S}$ is shown in Figs. 1 and 2. ${\rm Im}V$ is determined by Eqs. (6), (7), (23) and (24), ${\rm Re}V$ is the parameter. Solid and dot-dashed curves correspond to $\mid {\rm Re}V \mid=\Delta \Gamma /2$ and ${\rm Re}V=0$, respectively. Dashed curve corresponds to the copper plate and $V$ defined from (25). The amplitudes $f$ and $\bar{f}$ are taken from [17]. In the case ${\rm Re}V=0$ only the regeneration by absorption takes place. It is seen that ${\rm Re}V$ leads to the suppression of regeneration. Compared to [2], $\mid \!K_S(t)\!\mid ^2$ is about 10 times smaller. (Although the comparison with [2] is meaningless for the reasons given above.) 

\section{Conclusion}
The main results of this paper are given in the abstract. The most distinctive feature of the model presented above is the inverse $\Delta \Gamma $- and $\Delta m$-dependences of the amplitude of regenerated $K^0_{S}$, or parameter regeneration $r$ (see Eqs. (19)-(24) of Ref. [6]). The main uncertainty in the numerical results is conditioned by the uncertainty in the cross sections $\sigma (K^0N)$ and $\sigma (\bar{K}^0N)$. The same is also true for the previous results [1-5] since they have been obtained by means of above-mentioned cross sections as well. In this connection we would like to recall that $\Delta m$ is extracted from free-space oscillations without recourse to potentials of $K^0$ and $\bar{K}^0$. Nevertheless, in any case the regeneration should be described correctly. 

This paper accepted for publication in Chinese Physics C. The author is grateful to Michael Bayev for help in numerical calculations.


\newpage
\begin{thebibliography}{99}
\bibitem{1}
K.M. Case, Phys. Rev. {\bf 103}: 1449 (1956)
%
\bibitem{2}
M.L. Good, Phys. Rev. {\bf 106}: 591 (1957)
%
\bibitem{3}
M.L. Good, Phys. Rev. {\bf 110}: 550 (1958)
%
\bibitem{4}
T.D. Lee and C.S. Wu, Annu. Rev. Nucl. Sci. {\bf 16}: 511 (1966)
%
\bibitem{5}
E.D. Commins and P. H. Bucksbaum, {\em Weak Interactions of Leptons and Quarks}
(Cambridge University Press, 1983).
%
\bibitem{6}
V.I. Nazaruk, Int. J. Mod. Phys. E {\bf 25}: 1650104 (2016) 
%
\bibitem{7}
Xian-Wei Kang, Hai-Bo Li and Gong-Ru Lu, Phys. Rev. D {\bf 81}: 051901 (R) (2010) 
%
\bibitem{8}
X. W. Kang, H. B. Li, G. R. Lu, arXiv: hepph/1008.2845
%
\bibitem{9}
Yong-Feng Liu , Xian-Wei Kang , J. Phys. Conf. Ser. 738:  no.1, 012043 (2016)
%
\bibitem{10}
Xian-Wei Kang, Bastian Kubis, Christoph Hanhart et al, Phys.Rev. D {\bf 89}: 053015 (2014)
%
\bibitem{11}
V.I. Nazaruk, Int. J. Mod. Phys. E {\bf 26}: 1750007 (2017)
%
\bibitem{12}
V.I. Nazaruk, Eur. Phys. J. C {\bf 53}: 573 (2008)
%
\bibitem{13}
V.I. Nazaruk, Phys. Lett. B {\bf 337}: 328 (1994)
%
\bibitem{14}
V.I. Nazaruk, Int. J. Mod. Phys. E {\bf 21}: 1250056 (2012) 
%
\bibitem{15}
V.I. Nazaruk, Int. J. Mod. Phys. E {\bf 20}: 2377 (2011)
%
\bibitem{16}
Particle Date Group, Phys. Rev. D {\bf 66}: 010001 (2002)
%
\bibitem{17}
A.Di Domenico, Nucl. Phys. B {\bf 450}: 293 (1995)

\end{thebibliography}
\end{document}